# Discriminating between lepton number violating scalars using events with four and three charged leptons at the LHC

Francisco del Águila,[1] Mikael Chala,[1] Arcadi Santamaria,[2] and Jose Wudka[3]

[1]*CAFPE and Departamento de Física Teórica y del Cosmos,*
*Universidad de Granada, E–18071 Granada, Spain*

[2]*Departament de Física Teòrica, Universitat de València and IFIC,*
*Universitat de València-CSIC, Dr. Moliner 50,*
*E-46100 Burjassot (València), Spain*

[3]*Department of Physics and Astronomy,*
*University of California, Riverside CA 92521-0413, USA*

## Abstract

Many Standard Model extensions predict doubly-charged scalars; in particular, all models with resonances in charged lepton-pair channels with non-vanishing lepton number; if these are pair produced at the LHC, the observation of their decay into $l^\pm l^\pm W^\mp W^\mp$ will be necessary in order to establish their lepton-number violating character, which is generally not straightforward. Nonetheless, the analysis of events containing four charged leptons (including scalar decays into one or two taus as well as into $W$ bosons) makes it possible to determine whether the doubly-charged excitation belongs to a multiplet with weak isospin $T = 0$, $1/2$, $1$, $3/2$ or $2$ (assuming there are no excitations with charge $> 2$); though discriminating between the isosinglet and isodoublet cases is possible only if charged-current events cannot produce the doubly-charged isosinglet.

With the discovery of the Higgs boson at the LHC [1, 2] the minimal Standard Model (SM) appears to describe up to very high accuracy all phenomena with scales below a TeV. Still, neutrinos have a mass and hence, new degrees of freedom must be added to the SM; either their right-handed (RH) counterparts to allow sufficiently small Dirac masses, or heavy fields to induce the Weinberg operator [3], $\mathcal{O}^{(5)} = (\overline{L_L^c}\tilde{\phi}^*)(\tilde{\phi}^\dagger L_L)$, resulting in sub-eV Majorana neutrino masses after electroweak symmetry breaking: $\mathcal{L}_{m_\nu}^{(5)} = -x_{ij}\mathcal{O}_{ij}^{(5)}/M \to -(x_{ij}v^2/2M)\overline{\nu_{Li}^c}\nu_{Lj}$ with $v/\sqrt{2} = \langle\phi^0\rangle \sim 174$ GeV the Higgs vacuum expectation value (VEV). [1] In this second case lepton number (LN) is broken because $\mathcal{O}^{(5)}$ involves two lepton doublets with total LN equal to 2, and two SM Higgs doublets with vanishing LN. This breaking is, however, very small because its coefficient $xv/M = m_\nu/v$ is $\sim 10^{-12}$ to account for the observed neutrino masses, $m_\nu \sim 0.1$ eV. This slight breaking can indicate, for example, new physics (NP) at a very high scale $M \sim 10^{14}$ GeV, if $x \sim 1$, or at the TeV scale if $x \sim 10^{-11}$. This last example is the scenario we are a priori interested in: new particles near the TeV with LN violating (LNV) decays eventually observable at the LHC.

The simplest possibility results from the addition to the SM of a number of heavy RH Majorana neutrinos $N$, which has been extensively studied [4–7]. However, the corresponding LNV signal $l^\pm l^\pm W^\mp$ (with the $W$ decaying to jets and the background event carrying the missing charge) does not allow for searches in a broad range of heavy neutrino masses because heavy neutrino production is suppressed by mixing angles [8], and the final state does not allow for a very efficient resonance reconstruction since only one of the two same-sign leptons comes from the $N$ decay. The next simplest alternatives require the addition [2] of a scalar weak triplet $\Delta$ with hypercharge $Y_\Delta = 1$ (the electric charge $Q$ equals $T_3 + Y$, with $T_3$ the third component of isospin), or of a fermion isotriplet of 0 hypercharge; these correspond to the so-called *see-saw* models of type II and III, respectively, in contrast with type I when the heavy mediator is the neutrino $N$. An essential difference in the triplet cases is that they contain new particles that are pair produced with electroweak strength, since they are charged, and they can be easily reconstructed through their leptonic decays.

---

[1] $L_L^c = (\nu_L^c, l_L^c)$ is the SM lepton doublet with charge conjugated fields, $\psi_L^c = (\psi_L)^c = C\overline{\psi_L}^T$, $\psi_R^c = (\psi_R)^c = C\overline{\psi_R}^T$, and $\phi = (\phi^+, \phi^0)$ the SM Higgs doublet, with $\tilde{\phi} = i\sigma_2\phi^*$ and $\sigma_2$ the second Pauli matrix. In the text we write down column doublets in a row for convenience, when no confusion is expected.

[2] In the following it should be understood that when we refer to an "addition" of certain particles or to an "extension" containing such particles, we mean the extension of the SM that is obtained by adding the corresponding fields.

The corresponding signals have been also widely studied in the literature [9–13].

In this note we discuss the extension of the searches based on the see-saw of type II, generalizing the phenomenological approach proposed in Refs. [14–16]. In contrast with the see-saw models of type I and III, that involve the addition of heavy leptons, and whose decays involve at least three light fermions, the see-saw of type II results from the addition of a triplet of heavy scalars that couples to two light leptons with the same leptonic charge, and this will exhibit a resonant behavior in the corresponding di-lepton channel with |LN|=2. CMS [17] and ATLAS [18] have already set stringent bounds on this scenario, excluding scalar masses $\sim 400$ GeV with an integrated luminosity of $\sim 5$ fb$^{-1}$ at a center of mass energy of $\sqrt{s} = 7$ TeV, although with the assumption that the doubly-charged scalar component $\Delta^{\pm\pm}$ only decays into two same-sign leptons of the first two families. These limits are much less stringent when $\Delta^{\pm\pm}$ mainly decays into tau leptons, and even weaker if they have an appreciable branching ratio into $W^\pm W^\pm$; on the other hand, the reach is so high because it also benefits of rather small SM backgrounds. At any rate, the outstanding performance of these experiments will allow them to probe quite high masses at the LHC in the near future. Once a doubly-charged resonance is observed in a same-sign di-lepton channel, our purpose is then to discuss to what extent it can be established that it is a member of the scalar triplet mediating the see-saw of type II, or whether it belongs to another multiplet. In order to address this question we will first classify the type of scalar multiplet additions H containing such a resonance; these are characterized by their isospin $T$ and hypercharge $Y$, which in turn determine the leading contribution to the total cross sections for doubly-charged scalar pair production and, if not a weak singlet, for the associated production with its singly-charged partner. This will eventually allow the measurement of both the $T$ and $Y$ quantum numbers of the new multiplet H.

Any scalar multiplet containing a doubly-charged field can be coupled to like-charge di-leptons in a gauge invariant way by considering effective operators of high enough dimension. If we want to classify the scalar resonances coupling to lepton pairs with |LN| = 2 and observable (though not exclusively) in the doubly-charged channel with two charged leptons of the first two families, which will be the trigger for all these searches, we have to consider all invariant effective operators constructed with one of the two lepton bilinears with non-vanishing scalar couplings, $\overline{L_L^c}L_L, \overline{l_R^c}l_R$, [3] any number of SM Higgs doublets $\phi$ and the new scalar multiplet H they belong to. There are two such operators of dimension 4 (containing no SM Higgs doublets) generating the desired di-leptonic scalar decays: in one the new scalars couple to the bilinear with left-handed (LH) lepton doublets and in the other to the bilinear with RH lepton singlets. In the former case the new scalar multiplet will be a triplet $\Delta = (\Delta^{++}, \Delta^+, \Delta^0)$ with hypercharge 1 mediating the much-studied see-saw of type II, in the latter case the new scalar will be an isosinglet $\kappa^{++}$ with hypercharge 2. (A realistic model of this type has been discussed in some detail in Ref. [19].) Except for higher-order corrections to these renormalizable couplings, the couplings to fermions of other multiplets with doubly-charged components do not occur in renormalizable theories, but must involve higher-dimensional operators. [4] If this is the case, the fundamental theory must contain additional heavier fields that upon integration give rise to the higher order operators (couplings) we will now classify. [5] There are three independent operators of dimension 5 (with one SM Higgs doublet): there is a single operator involving a quadruplet $\Sigma = (\Sigma^{++}, \Sigma^+, \Sigma^0, \Sigma'^-)$ of hypercharge 1/2 coupling to two LH lepton doublets; and there are two independent operators involving a doublet $\chi = (\chi^{++}, \chi^+)$ of hypercharge 3/2 coupling to both |LN| = 2 lepton bilinears; [6] the model resulting from the addition of this doublet was briefly discussed in Ref. [14], and studied recently in Refs. [15, 16, 22]. Finally, there is one operator of dimension 6 (containing two SM Higgs doublets) coupling a quintuplet $\Omega = (\Omega^{++}, \Omega^+, \Omega^0, \Omega^-, \Omega^{--})$ of hypercharge 0 to two LH lepton doublets; this field can be assumed to be real, as we do in the following. There are other operators involving quintuplets of non-zero hypercharge but these also include scalars with electric charges larger than 2;

---

[3] The other combination $\overline{L_L^c}l_R$ requires a $\gamma^\mu$ insertion because of the fermions' chirality, and hence the presence of a covariant derivative to ensure the operator is Lorentz invariant; through use of integration by parts and the equations of motion the corresponding operators are then seen to be equivalent to the ones considered here.

[4] This is the reason why they did not appear in the listing in [20].

[5] Typically a heavier triplet coupling to the LH lepton bilinear, with a small VEV or/and with scalar couplings softly breaking LN.

[6] We can also write the Weinberg operator but with one of the Higgs doublets replaced by a new heavy doublet of hypercharge 1/2, [21] but this does not contain a doubly-charged component and cannot resonate in same-sign di-lepton channels. Moreover, there is only one combination of the lepton doublets that gives a triplet involving the same-sign charged lepton bilinear; any other possible invariant operator of that type must be related. Indeed, $(\widetilde{\overline{L}}_L\chi)(\phi^\dagger L_L) = (-1)^a(\widetilde{\overline{L}}_L\tau^a L_L)(\phi^\dagger \tau^{-a}\chi) + \frac{1}{2}(\widetilde{\overline{L}}_L L_L)(\phi^\dagger \chi)$, where the last operator does not include the doubly-charged bilinear combination.

| | | |
|---|---|---|
| Dimension 4 | Triplet $\Delta$ | $(\overline{\tilde{L}_L}\tau^a L_L)\Delta^{-a}$ |
| | Singlet $\kappa$ | $\overline{l_R^c}l_R\kappa$ |
| Dimension 5 | Quadruplet $\Sigma$ | $(-1)^{\frac{1}{2}-b}\mathrm{C}_{a,b}^{1\times\frac{1}{2}\to\frac{3}{2}}(\overline{\tilde{L}_L}\tau^a L_L)\phi^b\Sigma^{-a-b}$ |
| | Doublet $\chi$ | $(-1)^{1-a}(\overline{\tilde{L}_L}\tau^a L_L)(\phi^\dagger\tau^{-a}\chi),\ \overline{l_R^c}l_R(\tilde{\phi}^\dagger\chi)$ |
| Dimension 6 | Quintuplet $\Omega$ | $\mathrm{C}_{a,b}^{1\times 1\to 2}(\overline{\tilde{L}_L}\tau^a L_L)(\tilde{\phi}^\dagger\tau^b\phi)\Omega^{-a-b}$ |

Table I: Lowest-dimension, independent, gauge invariant effective operators coupling scalar multiplets H in the text with charges $|Q|\leq 2$ to lepton pairs with $|LN|=2$, observable in same-sign di-lepton channels. $\tilde{L}_L = i\sigma_2 L_L^c$, $\mathrm{C}_{m_1,m_2}^{j_1\times j_2\to j}$ are the corresponding Clebsch-Gordan coefficients and $\tau^a$ the Pauli matrices in the spherical basis, $A^{+1} = -\frac{1}{\sqrt{2}}(A_1 - iA_2), A^0 = A_3, A^{-1} = \frac{1}{\sqrt{2}}(A_1 + iA_2)$, times $\mathrm{C}_{a,-a}^{1\times 1\to 0}$, up to a global factor and sign: $\tau^{\pm 1} = \pm(\sigma_1 \mp i\sigma_2)/2,\ \tau^0 = \sigma_3/\sqrt{2}$.

the same is true of multiplets with isospin $> 2$. [7] We collect the corresponding operators in Table I.

In the following we will concentrate on the additions to the SM listed in Table I which are those with the lowest isospin and requiring lower dimension effective operators. They also exhaust the models with doubly-charged scalars resonating in same-sign di-lepton channels but without scalars with larger electric charges, that is, these multiplets satisfy

$$T + Y = 2\,. \tag{1}$$

The explicit expressions of the operators in Table I will not be needed in the simulations performed in this note. Indeed, the decay of a scalar into two leptons can be in general parametrized (after spontaneous symmetry breaking) by two independent couplings corresponding to the two fermion chiralities:

$$\overline{\psi_L^c}\psi_L' \mathrm{H}^{++}\,, \quad \text{and} \quad \overline{\psi_R^c}\psi_R' \mathrm{H}^{++}\,. \tag{2}$$

Thus, all operators containing $\overline{L_{Li}^c}L_{Lj}$ ($\overline{l_{Ri}^c}l_{Rj}$) reduce to the first (second) times a small (in general, flavor-dependent) coupling constant. Since the helicity of the final leptons cannot be measured, except eventually for the tau lepton [25], then although the operators in Table I involve definite fermion chiralities, we could use any of the two couplings above in the simulation. In the $\tau$ case, the dependence of the total cross section on the $\tau$ helicity due to the experimental cuts is anyway negligible. On the other hand, although the relative strength of the lepton couplings to different components within a given scalar multiplet is fixed by gauge invariance (inherent to the explicit form of the operators), we will not use these relations either in this letter. In a companion paper we will provide the missing details, including also the Feynman rules, the details of the applied cuts and of the event reconstruction procedures we will refer to later, as well as the discussion of other related signatures not studied here; we shall also provide the code we used to perform the corresponding simulations.

---

[7] Scalars with larger electric charges have a rich variety of striking decays, especially if they are mass degenerate with their doubly-charged partners [23, 24]. But these models are in general also more complicated.

Two remarks are in order to justify the way we proceed: (i) *LNV doubly-charged scalars are expected to decay slowly.* Although present limits on $\Delta^{\pm\pm}$ look quite stringent, the physics involved is quite rare. in the sense that LN breaking is very small, as already emphasized. This necessarily translates into very small decay widths into lepton pairs and/or gauge boson pairs (implying displaced vertices [10]) because both channels have different LN and their product must reflect the fact that LNV amplitude is minuscule, $m_\nu/v \sim 10^{-12}$. Similarly, the stringent limits on lepton flavor violation also strongly restricts the decay width at least into some lepton pairs. Hence, it also makes sense in general to look for decays into leptons that are as slow as into gauge bosons, which appreciably reduces present $\Delta^{\pm\pm}$ mass limits as stressed above. [8] (ii) *The effective Lagrangian approach is the appropriate way of describing these extension of the SM.* Indeed, specific models may not look simple because they have to explain these small numbers, which can be the (joint) result of small couplings, of quantum loop suppression factors, or of multiple layers of NP. So, in this context, once the SM has been largely confirmed, also including the Higgs sector, as well as the gap between the electroweak scale and the scale of NP, we have to consider all higher order effective operators coupling the new scalars H to like-charge di-leptons, although they may be suppressed by several powers of a heavy scale, as long as they are dominant.

Let us now discuss how to discriminate among the different H additions by exploiting the implications of gauge invariance, which completely fixes the scalar production cross sections. As a matter of fact, these cross sections are in general not only of electroweak size but are

---

[8] In the see-saw of type II if both channels have the same partial decay width, which implies $\langle \Delta^0 \rangle \sim 5 \times 10^{-5}$ GeV for $m_\Delta = 500$ GeV, then the decay length is $\sim 10\mu$m; see Ref. [11] for the explicit decay width expressions and further discussion. In general, for doubly-charged scalar masses heavier than few hundreds of GeV, $\left[\sum_{i,j=e,\mu,\tau} \Gamma(\Delta \to l_i l_j)\Gamma(\Delta \to WW)\right]^{-1} \approx 10^{-15} \left[\sum_{i=1,2,3} m_{\nu_i}^2\right]^{-1} \frac{m_W^4}{m_\Delta^4} > 10^5 \frac{m_W^4}{m_\Delta^4} \mu\text{m}^2$.

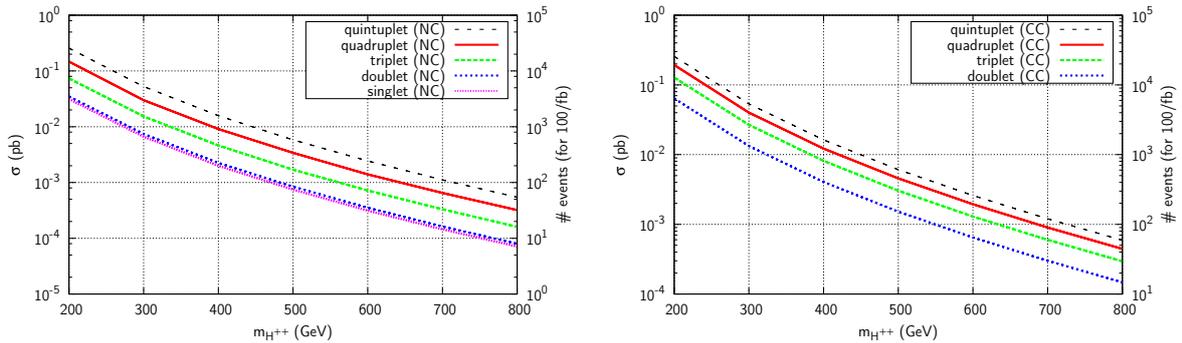

Figure 1: Doubly-charged scalar pair (left) and single (right) production at LHC for $\sqrt{s} = 14$ TeV, with scalars H belonging to a real quintuplet $\Omega$, a quadruplet $\Sigma$, a triplet $\Delta$, a doublet $\chi$ or a singlet $\kappa$ with hypercharge $Y = 0, 1/2, 1, 3/2$ and $2$, respectively. For a complex quintuplet the cross sections are double because there are two doubly-charged scalars, obviously, if mass degenerate. On the other hand, the singlet has no single production of doubly-charged scalars because it does not have a singly-charged component.

generated by the same types of couplings independently of the multiplet H, and differ only in the coupling strength, which only depends on the hypercharge and isospin of the multiplet. In our case, the doubly-charged scalar is either pair produced or produced in association with its singly-charged partner through the gauge couplings:

$$\mathcal{L}_\gamma^{H^{\pm\pm}} = ieQ(\partial^\mu H^{--})H^{++}A_\mu + \text{h.c.}\,, \qquad (3)$$

$$\mathcal{L}_Z^{H^{\pm\pm}} = \frac{ig}{c_W}(T_3 - Qs_W^2)(\partial^\mu H^{--})H^{++}Z_\mu + \text{h.c.}\,, \qquad (4)$$

$$\mathcal{L}_W^{H^{\pm\pm}} = \frac{ig}{\sqrt{2}}\sqrt{(T - T_3 + 1)(T + T_3)}[H^{++}(\partial^\mu H^-) - (\partial^\mu H^{++})H^-]W_\mu^- + \text{h.c.}\,, \qquad (5)$$

where $Q = 2$ and $T_3 = Q - Y$, and $c_W$ ($s_W$) the cosine (sine) of the Weinberg angle. Thus, measuring the total pair production cross section (generated by the $Z/\gamma$ couplings) we can determine the hypercharge, and hence $T_3$; this, together with the total associated production cross section (generated by the $W$ coupling), will allow an estimate of $T$. In Fig. 1 we plot the corresponding cross sections for the five cases listed in Table I. [9] It is important to note that since the hypercharge and the isospin are related by Eq. (1), it is

---

[9] Higher order contributions to doubly-charged scalar pair production as, for instance, those generated by vector boson fusion may need to be eventually considered for larger doubly-charged scalar masses. But in this case we can use the extra quarks in the forward direction to isolate this new type of events. The

enough to measure one of the two production cross sections in order to discriminate among the different possibilities. Both, neutral and charged production grow with the isospin, but pair production provides a better discriminator because it has smaller backgrounds and a smaller number of channels that contribute to the final modes of interest. Hence, in the following we will concentrate on this case. The growth of the cross sections mediated by neutral currents can be traced back to the different $Z$ couplings in Eq. (4), which are not so different in strength from the photon one. Although they also involve the quark couplings, which when properly taken into account make, for instance, the singlet and doublet cases indistinguishable. Thus, one must also rely on the absence of the associated production of a doubly and a singly-charged scalar in the singlet case to discriminate between them.

Any final mode requires not only the production but the subsequent decay of the new scalars. So, no particular channel allows the determination of the strength of the couplings involved in the production, but only their product by the corresponding branching ratios. Hence, although the production cross sections are fixed by gauge symmetry, we have to rely on measuring several (preferably all) decay channels in order to estimate the total cross section, and determine which of the scalar multiplets is being produced. Obviously, the scalar production in all cases is kinematically identical, except for the charge distribution which is related to the quark parton distribution functions. On the other hand, one can not rely on possible differences in the di-lepton branching ratios or in the kinematical distributions of the decay, because the former are model-dependent and the coupling constants multiplying the different operators in Table I can always be arranged to fit the observed number of events in a given final state; while the only kinematical observable which can be used to distinguish between operators is the tau lepton helicity, and taus do not need to be the most frequent decay product.

There is one last and essential point to discuss further before illustrating how to dis-

---

vector boson fusion cross section also depends on $T_3$ and $T$ only, and the analysis below goes through accordingly (we will provide the details in a forthoming publication). Single doubly-charged production through vector boson fusion may be only dominant for unusual models evading the stringent electroweak constraints on the VEV of the neutral scalar partner of the doubly-charged scalar boson [26]. In this case the doubly-charged scalar only decays into $W$ pairs, and does not resonate in the di-leptonic channel. We also assume that the SM Higgs does not have a large coupling to doubly-charged scalar pairs, and thus, that its contribution to doubly-charged pair production is negligible. Pair production through mixing with other scalars [27] is not considered either.

tinguish among the different H multiplets. In general, these scalars also decay into gauge bosons, in particular, into $W^\pm W^\pm$ if the scalar is doubly-charged. However, the corresponding branching ratio can be larger, of the same order or negligible when compared to the di-leptonic one, as already discussed above (see footnote 8) for the triplet case. [10] This is so because LN must be broken if neutrino masses are Majorana, as we assume, and therefore these new scalars must exhibit both decay modes at some level: if they decay into two leptons with the same leptonic charge, their LN would be well-defined and different from zero by 2 units, and if they only decay into a pair of gauge bosons, their LN would be preserved and equal to 0; only by having both decay channels their LN is not well-defined and LN violated. This is in practice realized by ensuring a small LNV VEV, for instance, when a triplet is added to the SM, by requiring $\langle \Delta^0 \rangle \neq 0$, and then inducing through the kinetic term the coupling

$$g^2 \langle \Delta^0 \rangle W^{\mp\mu} W^{\mp}_\mu \Delta^{\pm\pm} \,. \qquad (6)$$

The models with multiplets without neutral components which can acquire a VEV must include mixing terms violating LN in the scalar potential in order to generate this coupling to some order [29]. As indicated above, LN is violated very weakly and this small number is in general proportional to $y\eta$, where $y$ is the effective di-lepton Yukawa coupling in Eq. (2) and $\eta$ is proportional to a LNV VEV (similarly to the triplet case in Eq. (6)), $\langle H^0 \rangle /v$, and/or to a small mixing angle, possibly times a loop suppression factor. (This product also enters in the amplitude for neutrino-less double beta decay, and this further restricts the models [19, 20].) The constraint $y\eta \ll 1$ encompasses all scenarios found in specific models: $\eta$ much larger than, of the same order as or much smaller than $y$. In the first case the new scalars decay mainly into gauge bosons and their signals do not emerge from the background because their invariant masses can not be efficiently reconstructed [30]. In the other two cases the analysis we proposed can be carried out. If both decay channels are comparable, LNV could be experimentally confirmed by observing $l^\pm l^\pm W^\mp W^\mp$ events (and/or $l^\pm l^\pm W^\mp Z$ if $H^{\pm\pm}$ is single produced). If only di-lepton channels are observable, LNV may not be established at the LHC but the type of scalar multiplet could be still determined.

---

[10] Experimental limits [17, 18] are in general given assuming that the doubly and singly-charged scalars only decay into leptons. As pointed out, we allow these scalars also to decay into gauge bosons. But we neglect mass splittings within multiplets due to mixing with other scalars. If not, decay chains must be also considered [28].

The proposed analysis is based on the refinement and extension of the search for a doubly-charged resonance decaying into two same-sign leptons at LHC. Present limits [17, 18] are obtained assuming fixed branching ratios to di-leptons only. But, in general, a 100 % reconstruction is only obtained by summing all decay modes: $1 = \sum_{a=ll,l\tau,\tau\tau,WW} z_a$, $z_a \equiv \mathrm{Br}(\mathrm{H} \to a)$. Hence, as already emphasized, the total $\mathrm{H}^{\pm\pm}$ production cross section cannot be inferred from the observation of a single decay mode. However, one can try to measure it accounting for different channels. Indeed, the four charged lepton ($e$ or $\mu$) cross section for any given channel $ab$ can be written $\sigma_{ab} = (2 - \delta_{ab})\sigma z_a z_b$, where $\sigma$ is the total scalar pair production cross section we want to measure and $z_{a,b}$ the corresponding branching ratios, which in general include cascades into two leptons plus missing energy; note that we are dealing with extremely narrow resonances. Thus, the doubly-charged pair production cross section with both scalars decaying into two leptons of the first two families reads $\sigma z_{ll}^2$; whereas, for instance, the doubly-charged pair production cross section with one scalar decaying into two leptons of the first two families and the other to anything giving two charged leptons of the first two families, too, plus missing energy is written $\sigma_{lllp_T^{miss}} = \sigma_{llll} + 2\sum_{a=l\tau,\tau\tau,WW} \sigma z_{ll} z_a \mathrm{Br}(a \to ll + p_T^{miss})$. Hence, if we are able to reconstruct and estimate all $\sigma_{lla} \equiv 2\sigma z_{ll} z_a, a \neq ll$, besides $\sigma_{llll}$, we can then evaluate

$$\sigma = \left(\sigma_{llll} + \frac{1}{2}\sum_{a \neq ll} \sigma_{lla}\right)^2 / \sigma_{llll}. \qquad (7)$$

In the following we argue that this is feasible, knowing that experimentalists will easily improve on the assumptions being made here, especially when using real data. Assuming a heavy scalar mass $m_{\mathrm{H}^{\pm\pm}} = 500$ GeV, [11] doubly-charged scalar pairs are generated at LHC for a center of mass energy $\sqrt{s} = 14$ TeV using MADGRAPH5 [31], after implementing Eqs. (3–5), and the CTEQ6L1 parton distribution functions. Backgrounds are obtained with ALPGENV2.13 [32], whereas parton radiation and fragmentation are simulated with PYTHIAV6 [33] and the detector with DELPHESV1.9 [34]; details including sample selection and standard cuts will be presented in the companion paper. We then choose events with four charged leptons of the first two families and zero total charge, also requiring that one same-sign pair has an invariant mass within the interval $m_{\mathrm{H}^{\pm\pm}} \pm 40$ GeV. We find that the number of background events is $\sim 50$ for an integrated luminosity of 300 fb$^{-1}$, and that the number of signal events also depends on the multiplet the doubly-charged scalar belongs to, and on the assumed branching ratios $z_a$. In Table II we gather 4 different cases for illustration: $(z_{ll}, z_{l\tau}, z_{\tau\tau} + z_{WW}) = (1, 0, 0), (1/2, 1/2, 0), (1/2, 0, 1/2), (1/3, 1/3, 1/3)$, [12] for each multiplet addition. We specify in each case the total number of events passing the cuts, $(l^\pm l^\pm) l^\mp l^\mp p_T^{miss}$, and also have both like-charge pairs reconstructing the doubly-charged scalar mass, $(l^\pm l^\pm)(l^\mp l^\mp)$. We sum $\tau\tau$ and $WW$ events because we can easily disentangle the $\tau\tau + WW$ sample (with a very similar efficiency for both types of events) from the $ll$ and $l\tau$ ones, while distinguishing between both subsamples requires more sophisticated techniques. If we are interested in establishing the LNV decay $l^{\pm}l^{\pm}W^{\mp}W^{\mp}$ and measuring its cross section in this final mode, we obtain the best sensitivity by subtracting from the common sample those events consistent with the second same-sign lepton pair reconstructing two $\tau$ leptons with the doubly-charged scalar invariant mass. This is possible because there are 2 unknowns and 3 constraints when $p_T^{miss}$ is also measured. [13] In contrast, the $WW$ reconstruction cannot be done on an event-by-event basis because in this case there are 6 unknowns but only 5 constraints.

---

[11] Note that present limits are weakened when $z_{ll} + z_{l\tau}$ is appreciably smaller than 1.

|  |  | $(1, 0, 0)$ | $(\frac{1}{2}, \frac{1}{2}, 0)$ | $(\frac{1}{2}, 0, \frac{1}{2})$ | $(\frac{1}{3}, \frac{1}{3}, \frac{1}{3})$ |
|---|---|---|---|---|---|
| Quintuplet | $(l^\pm l^\pm) l^\mp l^\mp p_T^{miss}$ | $1307 \pm 38$ | $501 \pm 25$ | $362 \pm 22$ | $238 \pm 19$ |
|  | $(l^\pm l^\pm)(l^\mp l^\mp)$ | $1046 \pm 32$ | $261 \pm 16$ | $261 \pm 16$ | $116 \pm 11$ |
| Quadruplet | $(l^\pm l^\pm) l^\mp l^\mp p_T^{miss}$ | $765 \pm 30$ | $293 \pm 20$ | $212 \pm 18$ | $139 \pm 16$ |
|  | $(l^\pm l^\pm)(l^\mp l^\mp)$ | $612 \pm 24$ | $153 \pm 12$ | $153 \pm 12$ | $68 \pm 8$ |
| Triplet | $(l^\pm l^\pm) l^\mp l^\mp p_T^{miss}$ | $383 \pm 22$ | $147 \pm 16$ | $106 \pm 15$ | $70 \pm 13$ |
|  | $(l^\pm l^\pm)(l^\mp l^\mp)$ | $306 \pm 18$ | $77 \pm 9$ | $77 \pm 9$ | $34 \pm 6$ |
| Doublet | $(l^\pm l^\pm) l^\mp l^\mp p_T^{miss}$ | $189 \pm 17$ | $73 \pm 14$ | $53 \pm 13$ | $35 \pm 12$ |
|  | $(l^\pm l^\pm)(l^\mp l^\mp)$ | $151 \pm 12$ | $38 \pm 6$ | $38 \pm 6$ | $17 \pm 4$ |
| Singlet | $(l^\pm l^\pm) l^\mp l^\mp p_T^{miss}$ | $168 \pm 17$ | $64 \pm 13$ | $47 \pm 13$ | $31 \pm 12$ |
|  | $(l^\pm l^\pm)(l^\mp l^\mp)$ | $135 \pm 12$ | $34 \pm 6$ | $34 \pm 6$ | $15 \pm 4$ |

Table II: Number of expected signal events with four charged leptons, electrons or muons, at LHC with $\sqrt{s} = 14$ TeV and an integrated luminosity of 300 fb$^{-1}$ for a doubly-charged scalar mass of 500 GeV belonging to an electroweak quintuplet, quadruplet, triplet, doublet or singlet with hypercharge 0, 1/2, 1, 3/2 and 2, respectively, and different branching ratio $(z_{ll}, z_{l\tau}, z_{\tau\tau} + z_{WW})$ assumptions. After applying standard cuts, we require that two same-sign leptons reconstruct the scalar mass ±40 GeV and the other two reconstruct none, one or two taus, as well as the second doubly-charged scalar mass, or are compatible with its decay to $WW$. We also specify the number of events with the two same-sign pairs reconstructing both scalars. Only statistical errors are included.

---

[12] In definite models as in the see-saw of type II, the Yukawa couplings giving neutrinos a mass are the same mediating the like-charge di-leptonic scalar decay, and they are then constrained, [11] but this is not so in general.

Looking at Table II it is clear that the addition of a doublet and of a singlet cannot be distinguished in doubly-charged pair production in any channel (sum). Analogously, comparing the fourth column for the triplet (quadruplet) to the second one for the doublet (triplet) it is apparent that using only one channel we cannot always differentiate between the various multiplet extensions. But, as we have stressed before, counting the number of events in the three subsets, we *can* discriminate between the different scalar multiplets H, except between the doublet and the singlet. The three sub-samples are classified according to their kinematical properties (mainly the invariant mass of two same-sign leptons, $m_{ll}$, the missing momentum, $p_T^{miss}$, and the momentum fraction, $x$, carried out by the charged lepton in tau decays, as defined in footnote 13). Once the efficiency $\epsilon$ of the analysis for each sub-sample, $(l^{\pm}l^{\pm})(l^{\mp}l^{\mp})$, $(l^{\pm}l^{\pm})(l^{\mp}\tau^{\mp})$, $(l^{\pm}l^{\pm})(\tau^{\mp}\tau^{\mp} + W^{\mp}W^{\mp})$, is known, [14] the cross section for each subset, $\sigma_{llll,lll\tau,ll\tau\tau+llWW}$, and thus, the total cross section, can be obtained from real data. In Fig. 2 we plot the error estimate in the determination of $\sigma$ in Eq. (7) combining the three previous measurements for the quintuplet, quadruplet, triplet and doublet additions relative to the singlet one for an integrated luminosity of 100, 300 and 3000 fb$^{-1}$. It is apparent from this Figure and Table II that only a few hundred events are needed to distinguish between different multiplets. Hence, if a doubly-charged scalar is discovered, it will be possible to also decide which multiplet it belongs to by collecting enough statistics.

In order to distinguish the singlet from the doublet extensions we must look at the associated (charged) production, which is absent in the former case, and bounded from below in the latter one (assuming a signal is observed in pair production). It is then sufficient to search for three charged leptons plus missing energy, requiring two same-sign leptons to reconstruct the doubly-charged scalar, and requiring the other opposite-charged lepton and the missing momentum be compatible with a singly-charged scalar of similar mass. [15] This is also illustrated in Table III when assuming the same branching ratios for singly and doubly-charged scalars. Although when dealing with specific models one must calculate the different partial decay widths and sum them up to obtain the corresponding branching ratios for singly and doubly-charged scalars; for they are in general correlated.

---

[13] The momentum of an energetic $\tau$ can be assumed to align with the charged-lepton momentum it decays to: $xp_\tau^\mu = p_l^\mu$, with $0 < x < 1$.

[14] From our Monte Carlo simulations we estimate the sub-sample efficiencies including the corresponding branching ratios: $\epsilon_{lll} = 0.6$, $\epsilon_{lll\tau} = 0.09$ and $\epsilon_{ll\tau\tau} = \epsilon_{llWW} = 0.02$, respectively.

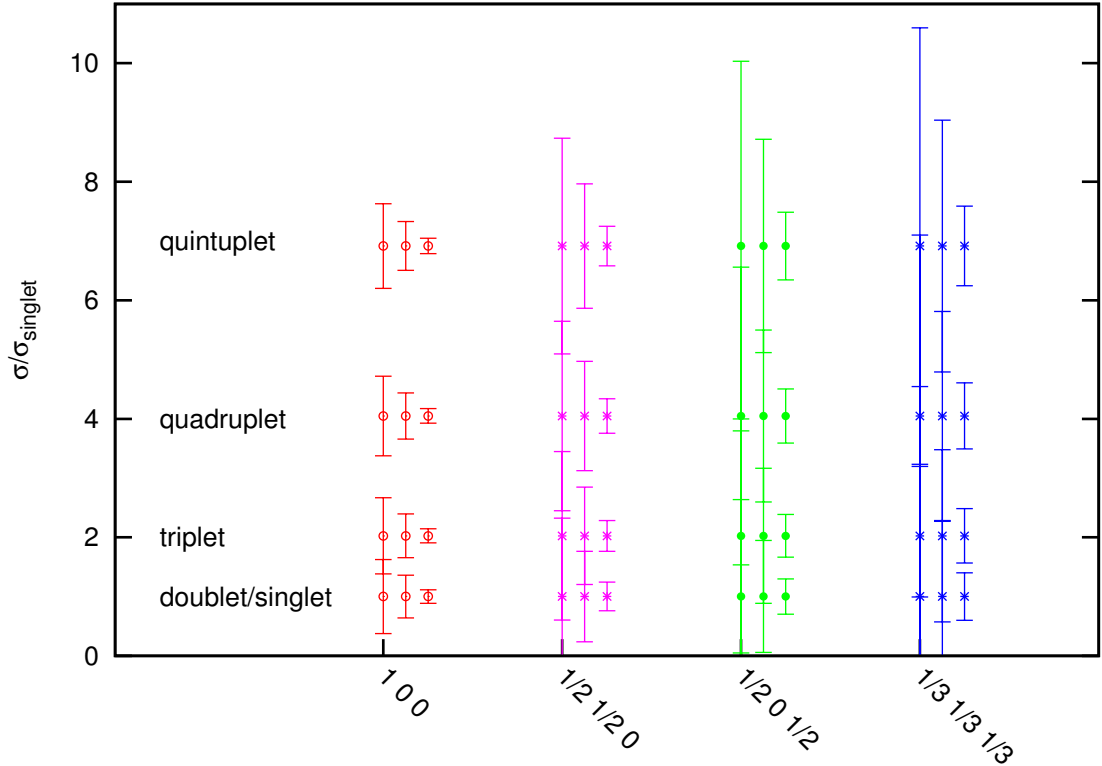

Figure 2: Error estimate for the measurement of the total cross section for doubly-charged scalar pair production in Eq. (7) for $m_{H^{\pm\pm}} = 500$ GeV at LHC with $\sqrt{s} = 14$ TeV and an integrated luminosity of 100, 300 and 3000 fb$^{-1}$ (from left to right), assuming it belongs to a weak quintuplet, quadruplet, triplet or doublet (singlet) with hypercharge 0, 1/2, 1 and 3/2 (2), respectively.

Finally, one can wonder about other final states; for instance, including semi-leptonic

---

[15] This test makes use of the distribution of the transverse mass of the opposite-sign lepton and $p_T^{miss}$, and of the number of jets.

| $(l^{\pm}l^{\pm})(l^{\mp}p_T^{miss})$ | $(1, 0, 0)$ | $(\frac{1}{2}, \frac{1}{2}, 0)$ | $(\frac{1}{2}, 0, \frac{1}{2})$ | $(\frac{1}{3}, \frac{1}{3}, \frac{1}{3})$ |
|---|---|---|---|---|
| Quintuplet | $1011 \pm 34$ | $283 \pm 21$ | $261 \pm 20$ | $130 \pm 17$ |
| Quadruplet | $592 \pm 27$ | $166 \pm 18$ | $153 \pm 17$ | $76 \pm 15$ |
| Triplet | $296 \pm 21$ | $83 \pm 15$ | $77 \pm 15$ | $38 \pm 14$ |
| Doublet | $146 \pm 17$ | $41 \pm 13$ | $38 \pm 14$ | $19 \pm 13$ |
| Singlet | $0 \pm 12$ | $0 \pm 12$ | $0 \pm 12$ | $0 \pm 12$ |

Table III: As in Table II but for for the produciton of a single doubly-charged scalar in association with a singly-charged scalar of a similar mass. We also require that the opposite sign lepton (electron or muon) and the missing momentum (corresponding to a neutrino) are compatible with the di-leptonic decay of the singly-charged scalar.

$WW$ decays for the second doubly-charged scalar. This signal, however, cannot be separated from the same final state from single production with the associated singly-charged scalar decaying into $WZ$; this and other examples allowing for consistent checks will be also studied in detail in the companion paper. Obviously, once the possibility of discriminating among different H multiplets is established, there will be many other cross-checks that will lead to alternative ways of discriminating among the scalar multiplet additions.

**Acknowledgements**

We thank J. Santiago and A. Aparici for useful discussions and the careful reading of the manuscript. This work has been supported in part by the Ministry of Economy and Competitiveness (MINECO), under the grant numbers FPA2006-05294, FPA2010-17915 and FPA2011-23897, by the Junta de Andalucía grants FQM 101 and FQM 6552, by the "Generalitat Valenciana" grant PROMETEO/2009/128, and by the U.S. Department of Energy grant No. DE-FG03-94ER40837. M.C. is supported by the MINECO under the FPU program.